\pdfoutput=1
\documentclass[aps,prl,twocolumn,amsmath,amssymb,superscriptaddress]{revtex4-1}

\usepackage{graphicx}
\usepackage{dcolumn}% Align table columns on decimal point
\usepackage{bm}% bold math
\usepackage[caption=false]{subfig} 
\usepackage{color}
\usepackage[titles,subfigure]{tocloft}
\usepackage{hyperref}
\usepackage[titletoc,toc,title]{appendix}

\newcommand{\beq}{\begin{eqnarray}}
\newcommand{\eeq}{\end{eqnarray}}

\newcommand{\bpm}{\begin{pmatrix}}
\newcommand{\epm}{\end{pmatrix}}
\newcommand{\om}{\omega}

\newcommand{\ket}[1]{| #1 \rangle}
\newcommand{\bra}[1]{\langle #1 |}
\newcommand{\dirac}[2]{\langle #1 | #2 \rangle}

\begin{document}

\title{Shift charge and spin photocurrents in Dirac surface states of topological insulator}

\author{Kun Woo Kim}
\affiliation{School of Physics, Korea Institute for Advanced Study, Seoul 02455, Korea}
\date{\today}
\author{Takahiro Morimoto}
\affiliation{Department of Physics, University of Californaia, Berkeley, CA 94720}
\date{\today}
\author{Naoto Nagaosa}
\affiliation{RIKEN Center for Emergent Matter Science (CEMS), Wako, Saitama, 351-0198, Japan}
\affiliation{Department of Applied Physics, The University of Tokyo, Tokyo, 113-8656, Japan}
\date{\today}

\begin{abstract}

The generation of photocurrent in condensed matter is of main interest for photovoltaic and optoelectonic applications. Shift current, a nonlinear photoresponse, has attracted recent intensive attention as a dominant player of bulk photovoltaic effect in ferroelectric materials. In three dimensional topological insulators $\text{Bi}_2\text{X}_3$ (X: Te, Se), we find that  Dirac surface states with a hexagonal warping term carry shift current by linearly polarized light. In addition, shift spin-current is introduced with the time-reversal symmetry breaking perturbation. The estimate for the magnitudes of the shift charge- and spin-currents are 0.13$I_0$ and 0.21$I_0$(nA/m) with the intensity of light $I_0$ measured in $(\text{W}/\text{m}^2)$, respectively, which can offer a useful method to generate these currents efficiently. 

\end{abstract}
\maketitle

{\it Introduction and background}-  Interaction of light with topological 
matters has drawn keen attention in condensed 
matter community. 
%Precise control of light in laboratories provides fruitful playground for 
%both theorists and experimentalists. 
Not only the characterization of topological matter 
has been done by light-assisted experimental tools that probe electronic 
bands~\cite{hsie2008topo,hsie2009obse2,xia2009obse} and its spin 
textures~\cite{hsie2009obse,hsie2009tuna}, optical conductivities, Kerr rotation~\cite{agui2012tera,jenk2010tera,tse2010gian,tse2011magn,shim2013quan}, Farady effect~\cite{sush2010far,tse2010gian,tse2011magn,shim2013quan}, 
etc, but also light  plays active roles to dynamically induce topological phase 
transitions~\cite{kita2011tran,lind2011floq,rech2013phot,usaj2014irra,
kata2013modu,dehg2014diss} and generates a new type of topological 
excitations~\cite{karz2015topo}.

Among them, photocurrent generation 
in condensed matters is an active research field closely related to light-harvesting 
energy research~\cite{choi2009swit,youn2012firs,youn2012firs2, heo2013effi, lind2014enha} 
and optoelectronic device applications~\cite{wund2010spin,bona2010grap,wang2012elec,
yu2013topo,sanc2014phot,cook2015desi}. 
{Particularly, shift current is the nonlinear optical responses, where 
the d.c. current is produced by light irradiation without the
external electric field or potential gradient, whose direction is determined
by the crystal structure or the direction of the electric polarization. 
It is suggested to play a dominant 
role in bulk photovoltaic effect of ferroelectric materials
including hybrid perovskite materials~~\cite{youn2012firs,youn2012firs2}. 

Further optimal engineering of materials to carry shift current is under 
studies~\cite{cook2015desi} to enhance photovoltaic efficiency.
The shift current is allowed in noncentrosymmetric systems since it is proportional 
to the square of the electric field $\bm{E}$. Specifically, it is of 
topological origin that is driven by the Berry phase connection $\bm{a}_n(\bm{k})$ of Bloch 
wavefunctions ( $n$: band index, $\bm{k}$: crystal momentum ) and is
allowed in noncentrosymmetric systems. This is because the Berry curvature 
$\bm{b}_n (\bm{k}) = \nabla_{\bm{k}} \times \bm{a}_n (\bm{k})$ 
satisfies $\bm{b}_n (\bm{k}) = - \bm{b}_n (-\bm{k})$ due to the time-reversal 
symmetry $\mathcal{T}$. The inversion symmetry $\mathcal{I}$ further gives 
the constraint $\bm{b}_n (\bm{k}) = - \bm{b}_n (\bm{k})$, which concludes
$\bm{b}_n (\bm{k}) = \bm{0}$ and one can choose the gauge 
$\bm{a}_n(\bm{k}) =\bm{0}$.}
{Therefore, it is essential to break either $\mathcal{T}$ or 
$\mathcal{I}$ for the nonzero $\bm{a}_n(\bm{k})$.
The physical meaning of the Berry connection $\bm{a}_n(\bm{k})$ is the 
intracell coordinates of the electron in the band $n$, which are related to 
the electric polarization $\bm{P}$ in ferroelectrics~\cite{rest1994macr,rest2007theo}.  
Therefore, the difference between $\bm{a}_{n=c}(\bm{k})$
and $\bm{a}_{n=v}(\bm{k})$ 
represents the shift of the electrons associated with the 
optical transitions from the valence band $v$ to the conduction band $c$. 
Under steady optical pumping, the constantly induced polarization described above results in the dc current, which is
the shift current.
  
Topological surface states are good candidates to support shift currents 
as the inversion symmetry $\mathcal{I}$ is always broken at the 
surface of materials. We show in this paper that  linearly polarized light can generate 
shift currents on warped topological surface states~\cite{chen2009expe,fu2009}  in
 $\text{Bi}_2\text{X}_3$-type topological insulators. Here the warping term breaks the 
mirror symmetry in one direction, and this determines unique combination of current and 
electric field directions to have nonzero shift currents. 
Furthermore, we propose in this paper the shift spin-current which is also of 
topological origin described by Berry phase connection. As for the surface states of topological 
insulators, we propose that pure shift spin-current (without shift charge-current)
is induced by breaking the time-reversal symmetry $\mathcal{T}$ with magnetic doping.
Its spin direction can be controlled by the direction of surface magnetization in topological 
insulators. Equipped with a shift charge- and spin-current production on Dirac surface 
states, improved platforms for opto-electronic and opto-spintronic devices can be provided 
by topological insulators. 

%This manuscript is organized in the following order. We first introduce the definition of shift 
%current for two-band system as a second order nonlinear response function with its 
%physical interpretaion. Then, the expression derived by \kun{Morimoto and 
%Nagaosa\cite{mori2015,mori2015shift}} from the Floquet formalism is shown to explain 
%in more general context that the shift current susceptilibity is a correlation function of 
%paramagnetic current and diamagnetic current. From this, the susceptilibity of spin-shift 
%current expression is naturally following. In subsequent sections, we show that the charge-shift 
%current is not allowed in inversion symmetric system, and the time-reversal symmetry 
%prohibits the spin-shift current. We suggest concrete and experimentally realizable 
%perturbations on Dirac surface states to induce those shift currents in topological insulators. 

{\it Model and formalism for shift charge- and spin-currents}- This nonlinear direct 
photocurrent can be understood in several ways as explained below. 
To discuss in concrete terms, let us consider a two-band system under a periodic time-dependent electric field with frequency $\Omega$, ${\vec{E}(t)}=-{\partial_t \vec{A}(t)} = -i\Omega (\vec{A}e^{i\Omega t}+c.c.)$. Treating the external 
field classically, we can write a Floquet Hamiltonian in the basis of valence and conduction bands~\cite{mori2015} (hereafter, $\hbar=1$ and $e=1$ is used):

\beq
H_F = \bpm \epsilon_v + \hbar \Omega & i\vec{A}^*\cdot \vec{v}_{vc} 
\\ -i\vec{A}\cdot \vec{v}_{cv}  & \epsilon_c \epm, \label{fH}
\eeq
where we concentrate on one valence band with a single photon absorbed and one conduction 
band.  $\epsilon_{c,v}$ are eigenvalues of original Hamiltonian $H$. Two bands in the Floquet 
Hamiltonian are coupled by time-dependent terms, $\bra{u_c} \frac{1}{T}\int_0^T H(\vec{k}-\vec{A}(t)) e^{i\Omega t } dt \ket{u_v} $. By taking   the linear order of external field only, the coupling can be expressed in terms of Fermi velocity: 
$-i \vec{A}\cdot \vec{v}_{cv}= -i\sum_{j} A_j v_{cv}^j$ with 
$v_{cv}^j=\bra{\psi_c} \partial_{k_j} H  \ket{\psi_v}= |v_{cv}^j| e^{i\varphi_{cv}^j}$. 
Sipe and Shkrebtii~\cite{sipe2000seco} obtained a simplified expression for a linearly polarized light case by computing photocurrent perturbatively, $J_j^{shift}=\sum_{i=x,y} \chi_j^{ii} E_i E_i$: 

\beq
\chi_{j}^{ii}
&=& \frac{\pi}{\Omega^2}\int \frac{d^d\vec{k}}{(2\pi)^d} \delta(\omega_{cv}-\Omega) 
|v_{vc}^i|^2 (\partial_{k_j} \varphi_{cv} ^i+ a_c ^j- a_v^j)  \nonumber \\
\label{shiftgeneral}  
\eeq
where $i$ and $j$ are cartesian coordinates,  and Berry connection 
$a_{c}^j = -i \dirac{\psi_c}{\partial_{k_j}\psi_c}$ and 
$a_v^j= -i \dirac{\psi_v}{\partial_{k_j}\psi_v}$ for conduction and valence band, respectively. 
The contributions are from the states satisfying the resonant condition: $\omega_{cv}-\Omega =\epsilon_c-\epsilon_v-\Omega$.  
The bracket on the right side is called a shift vector as the Berry connection $a_c^j$ indicates 
a shifting in real space of conduction band Bloch wave function along $j_{th}$ direction. 
$(a_c^j-a_v^j)$ is a difference of shifting between conduction and valence band, and 
$\partial_{k_j}\varphi_{cv}^i$ maintains the gauge invariance. As a result, a shift current is 
originated from the Bloch wave function's position change in real space due to electronic excitation 
from a valence band to a conduction band. 

An alternative derivation of the photocurrent expression was obtained by 
two of the authors previously~\cite{mori2015,mori2015shift} by
using Floquet formalism with Keldysh Green's function. They re-derived the three second-order photocurrents in easier manner compared to previous approaches~\cite{sipe2000seco,kral2000quat,kral2000quan}. In particular, the expression of shift current for a general gauge vector $\vec{A}$ is:
\beq
J_{j} =\int \frac{d^d{k}}{(2\pi)^d} \,\, \Im \left[ \delta(\omega_{cv}-\Omega) (\vec{A}^*\cdot \vec{v}_{vc}) \,  
 (\vec{A}\cdot \partial_{k_j}\vec{v})_{cv}  \right] \label{photocurrent}
\eeq
where 
\beq
(\vec{A}\cdot \partial_{k_j} \vec{v})_{cv} &=& \sum_i A_i \bra{u_c}\partial_{k_j} v^i\ket{u_v} 
\eeq
Provided that~\cite{mori2015} $(\partial_{k_j}v^i)_{cv}=\partial_{k_j}v_{cv} -\bra{\partial_{k_j} u_c}v\ket{u_v}-\bra{u_c}v\ket{\partial_{k_j} u_v}$, one can recover the shift current expression found by Sipe and Shkrebtii~\cite{sipe2000seco} in terms of Berry connections.
Expressed in terms of $v_{vc}^i$ and $(\partial_{k_j} v^l)_{cv}$, Eq.\eqref{photocurrent}  just looks like a correlation function between paramagnetic current and diamagnetic current. A Kubo conductivity in the linear response theory is obtained by taking a correlation function between two paramagnetic currents, $j^i_{para}=\left.\frac{\partial H(\vec{k}-\vec{A}(t))}{\partial A_i}\right|_{\vec{A}=0}=-v^i$. Eq.\eqref{photocurrent} is still a current-current correlation function, however, since a diamagnetic current contains a gauge vector, $j^l_{j,dia}=\left.\frac{\partial^2 H(\vec{k}-\vec{A}(t))}{\partial A_l \partial A_j} \right|_{\vec{A}=0} A_l =\left( \partial_{k_j} v^l \right)A_l$, the photocurrent $J$ is proportional to the square of electric field. 
Note that the three point correlation function of paramagnetic current also gives the second order photocurrent response as injection currents are obtained by Hosur~\cite{hosu2011}. In contrast, the susceptibility of current $J_j=\sum_{il}\chi_j^{il}E_iE_l$ in Eq.\eqref{photocurrent} can be computed by the following one-loop integration:

\beq
\chi_{j}^{il} =\frac{1}{\Omega^2}\int dk Tr [ v^i G(\om+\Omega,\vec{k}) 
(\partial_{k_j}v^l) G(\om,\vec{k})],  \label{suscep1} 
\eeq 
where $\int dk = \int d^d\vec{k} d\om/(2\pi)^{d}$ and $ G(\om,\vec{k}) = [\om-H]^{-1}$. We want to emphasize that the diamagnetic current, $\partial_{k_j} \hat{j}^l$, is responsible for the shifting of charge center, and it contains the index of photocurrent direction.

Lastly, we introduce a shift spin-current, which is an extension of a shift charge-current discussed so far. Having considered the shifting of charge center upon excitation from a valence to a conduction band, one can also think of the shifting of spin center that leads a shift spin-current. This can be most conveniently obtained from the one-loop diagram formulation: we take a particular spin direction, $\hat{s}_n$, for a diamagnetic current response, while we keep a paramagnetic current to be coupled to external electric field: 

% Explictly, to obtain this quantity,  in the current-current correlation function, \eqref{suscep1}, we can interpret one current operator as an external field perturbation and the other as a current response. To obtain a shift spin-current susceptibility, we pick up a particular spin direction by taking the anticommutation of a spin operator with either paramagnetic current or diamagnetic current. These two options depends on which term we will take as a perturbation and as a response. The total shift spin-current susceptibility is the summation of them: 
\beq
\chi_{j,\hat{s}_n}^{il} &=& \frac{1}{\Omega^2}\int dk  Tr [ v^i  G(\om+\Omega,\vec{k}) \{ \hat{s}_n, (\partial_{k_j}v^l) \} 
G(\om,\vec{k})] ,\nonumber \\ \label{spinshift}
\eeq 
where if $\hat{s}_n = \sigma_0/2$, this will recover the previous expression, Eq.\eqref{suscep1}. Young and Rappe~\cite{youn2013pred}  first discussed the spin bulk-photovolatic effect in antiferromagnets by computing shift current for spin-up and spin-down, independently. Our expression, Eq.\eqref{spinshift}, is applicable to more general cases with spin-orbit couplings.

%In the following, we explicitly show that a state and its inversion symmetry partner state makes exactly opposite contributions in both shift charge- and spin-currents. While, a state and its time-reversal symmetry partner state 

%Thus, a system with the broken inversion symmetry can carry shift currents. Then, we show that a shift spin-current is forbidden in systems with the time-reversal symmetry for a linearly polarized light. 

In the following, we explicitly show that the breaking of the inversion symmetry $\mathcal{I}$ is required to have a shift charge-current, and the breaking of the time-reversal symmetry $\mathcal{T}$ is  additionally required to have a shift spin-current.  To demonstrate, we consider two examples in 3d topological insulators: warped Dirac surface states carry shift charge-currents, and massive Dirac surface states with band-bending carry shift spin-currents.

{\it Shift charge current in Dirac surface states of topological insulators}- A system with the inversion symmetry carries a zero shift current. We first review this fact. This is intuitively correct, because a state $\ket{k}$ and its inversion symmetric partner $\mathcal{P}\ket{k}$ will be shifted in opposite direction after being excited to a conduction band. More formally, the inversion symmetry gives the following relation:

\begin{figure}
\begin{center}
\includegraphics[width=.9\columnwidth]{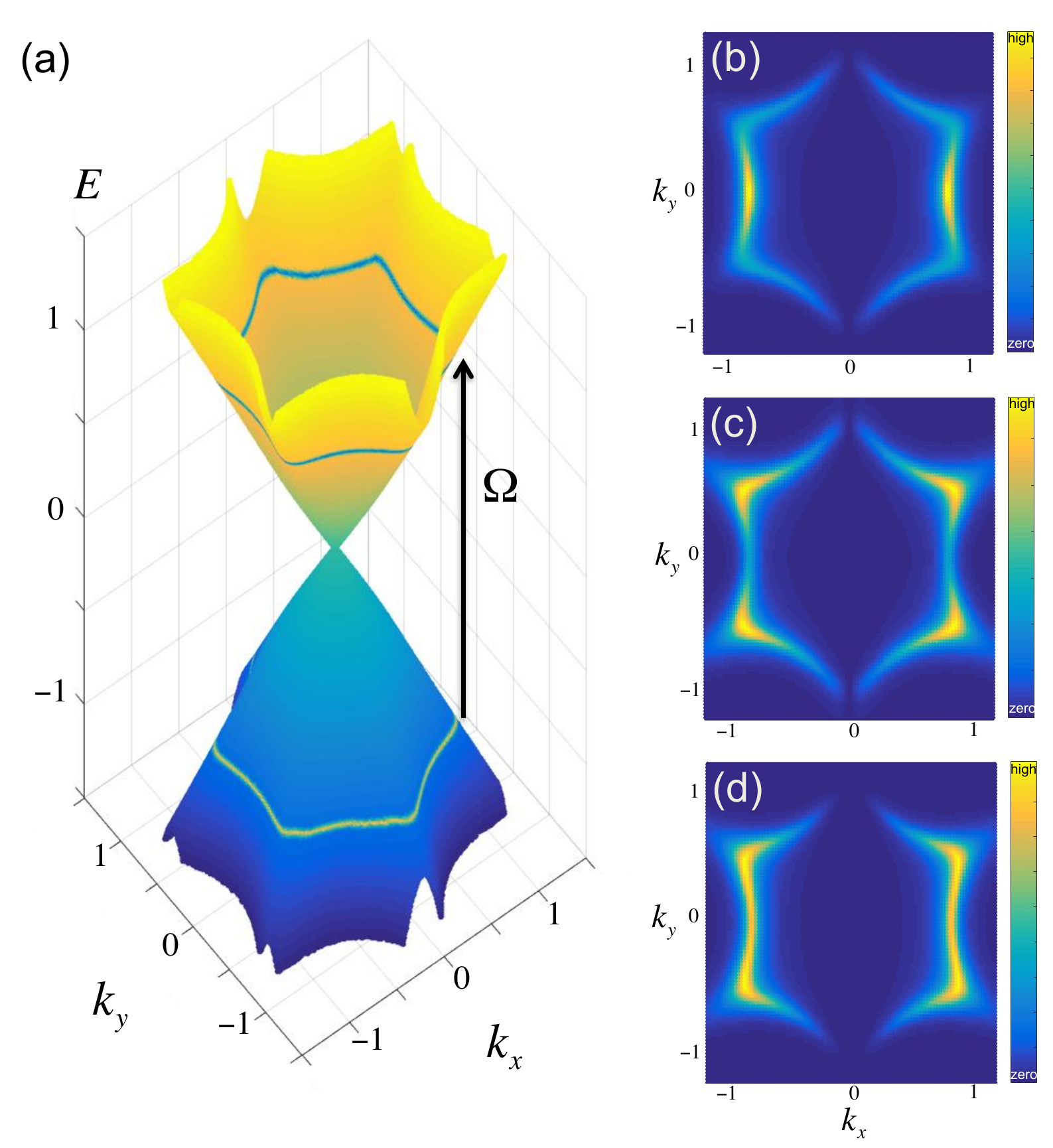}
\caption{ The susceptibility of shift charge-current, $J_{y,warp}=\chi_y^{yy}E_y^2$, is decomposed into the parts of perturbations and responses. (a) Warped Dirac surface states with resonance condition in distinct colors. ${k}_{x,y}$ and ${E}$ are in unit of $\sqrt{{v_F}/{\lambda}}$ and $\sqrt{{v_F^3}/{\lambda}}$, respectively. (b) The coupling strength of $\ket{\vec{k}}$  to a diamagnetic current response, $\delta(\om_{cv}-\Omega)|(\partial_{k_y}v^y)_{cv}|$, is plotted. (c) The coupling strength of  $\ket{\vec{k}}$ to an external electric field, $\delta(\om_{cv}-\Omega)|(v^y)_{vc}|$, is plotted. (d) The contribution of shift charge-current from  $\ket{\vec{k}}$, $\delta(\om_{cv}-\Omega)|(v^y)_{vc}(\partial_{k_y}v^y)_{cv}|$, is plotted. }
\end{center}
\end{figure}

\beq 
v_{vc}^i (\vec{k})\partial_{k_j} v_{cv}^l (\vec{k})
&=& (\partial_{k_i} H(\vec{k}) )_{vc} (\partial_{k_j} \partial_{k_l} H(\vec{k}) )_{cv},\nonumber \\
&\xrightarrow{\mathcal{I} }&  (-\partial_{k_i} H(-\vec{k}) )_{vc} (\partial_{k_j} \partial_{k_l} H(-\vec{k}) )_{cv},\nonumber \\
&=& -v_{vc}^i (-\vec{k})(\partial_{k_j} v^l (-\vec{k}))_{cv}.
\eeq
In the second equality, we use the inversion symmetry relation of Hamiltonian: $H(\vec{k}) = e^{-ikr}\hat{\mathcal{H}}e^{ikr} \xrightarrow{\mathcal{I} }  e^{-ikr}\mathcal{P}^{-1}\hat{\mathcal{H}} \mathcal{P} e^{ikr}= \mathcal{P}^{-1}H(-\vec{k}) \mathcal{P}  $ (see supplementary information). Therefore,
\beq
\chi_j^{il}= \frac{1}{2}\int && \frac{d^d\vec{k}}{(2\pi)^d} \delta (\om_{cv}-\Omega) \Im 
\bigg[ v_{vc}^i(\vec{k} )(\partial_{k_j} v^l (\vec{k}))_{cv}
\nonumber \\ && + v_{vc}^i(-\vec{k} )(\partial_{k_j} v^l (-\vec{k}))_{cv}\bigg] =0
\eeq
Note that, since the spin is invariant under the inversion symmetry operation, $\mathcal{I}: \hat{s}_n \rightarrow \hat{s}_n$, the susceptibility of shift spin-current is also exactly cancelled between inversion symmetry partners.

Now we study Dirac surface states in 3d topological insulators that are localized near open surfaces. Its low energy effective Hamiltonian is $H_0= -v_Fk_y\sigma_x +v_F k_x \sigma_y$, for which the inversion symmetry is absent. But, $H_0(\vec{k}-\vec{A}(t))$ does not contain diamagnetic terms ($\sim A_i A_l$), hence no shift current. With warping effects induced by $C_{3v}$ crystalline symmetry in discovered 3d topological insulators of $\text{Bi}_2\text{X}_3$ (where $\text{X}=\text{Te},\,\text{Se}$) type~\cite{chen2009expe}, we find that shift charge-currents are non-zero. The warped Dirac surface state Hamiltonian is~\cite{fu2009}:

\beq
H &=& -v_F k_y \sigma_x + v_F k_x \sigma_y +\lambda(k_x^3-3 k_xk_y^2)\sigma_z.\label{warpedham}
\eeq
This Hamiltonian is mirror symmetric along x, but not along y.
As a result, the susceptibility $\chi_j^{il}$ ($j,i,l=x,y$) can be non-zero when the number of index $y$ appears odd times so that that its inversion symmetry partner does not cancel the shift current. Indeed, a straightforward calculation of current-current correlation function Eq.\eqref{suscep1} shows that non-zero components 
are (see supplementary information):

\beq
\chi_y^{xx}=\chi_x^{yx}=-\chi_y^{yy} = \frac{3\lambda }{16v_F^2}
\eeq

Upon a linearly polarized light $\vec{E}(t) = E_0 (\cos \phi \hat{x}  + \sin\phi \hat{y} ) 
\cos \Omega t$, an induced shift currrent is:
\beq
\vec{J}_{warp} &=& -\frac{3\lambda}{16v_F^2} [ E_y^2 \hat{y} - E_x^2 \hat{y} - E_x E_y 
\hat{x} - E_y E_x \hat{x}]  \nonumber \\
&=& \frac{3 \lambda E_0^2 }{16v_F^2} [\sin2\phi \hat{x} + \cos2\phi \hat{y}]
\eeq
Interestingly, the shift charge-photocurrent found here is photon-freqeuncy independent. This is in a sharp contrast to the injection photocurrent calculation~\cite{junc2013} based on the Fermi's golden rule, where a photocurrent by linearly polarized light is a function of frequency $\Omega$.

Our numerical estimation is $J_{warp}=0.13 I_0(nA/m)$, where $I_0$ is the intensity of light (see supplementary information), using warping strength and Fermi velocity estimated by Fu~\cite{fu2009}.
The photocurrent value is comparable to the theoretical estimation of circular photogalvanic current (CPGC) by Hosur~\cite{hosu2011} where  photocurrent $0.1 I_0 (nA/m)$ is estimated for $10(T)$  in-plane magnetic field, which is to break the in-plane rotational symmetry. Note that the generation of CPGC is based on the selective excitation from a valence band to a conduction band, and this can be alternatively computed using the Fermi's golden rule~\cite{junc2013}.

In experiments~\cite{okad2016,mciv2012,olbr2014room}, photocurrents are measured in Dirac surface states using oblique incident light, which is again to break the in-plane rotational symmetry. Assuming the sample size as $1(mm^2)$, the photocurrent density is in the order of $1 I_0 (pA/m)$, which is two orders of magnitude lower than the theoretical  estimation of shift currents on warped Dirac surface states.

\textit{Shift spin-current in Dirac surface states of topological insulator with 
magnetic ordering}- A shift spin-current generated by one state is exactly cancelled by its time-reversal partner state. Thus, the absence of both inversion and
time-reversal symmetries is required for a system to carry
shift spin-currents. Specifically, due to the time-reversal operation,

\begin{figure}
\begin{center}
\includegraphics[width=.9\columnwidth]{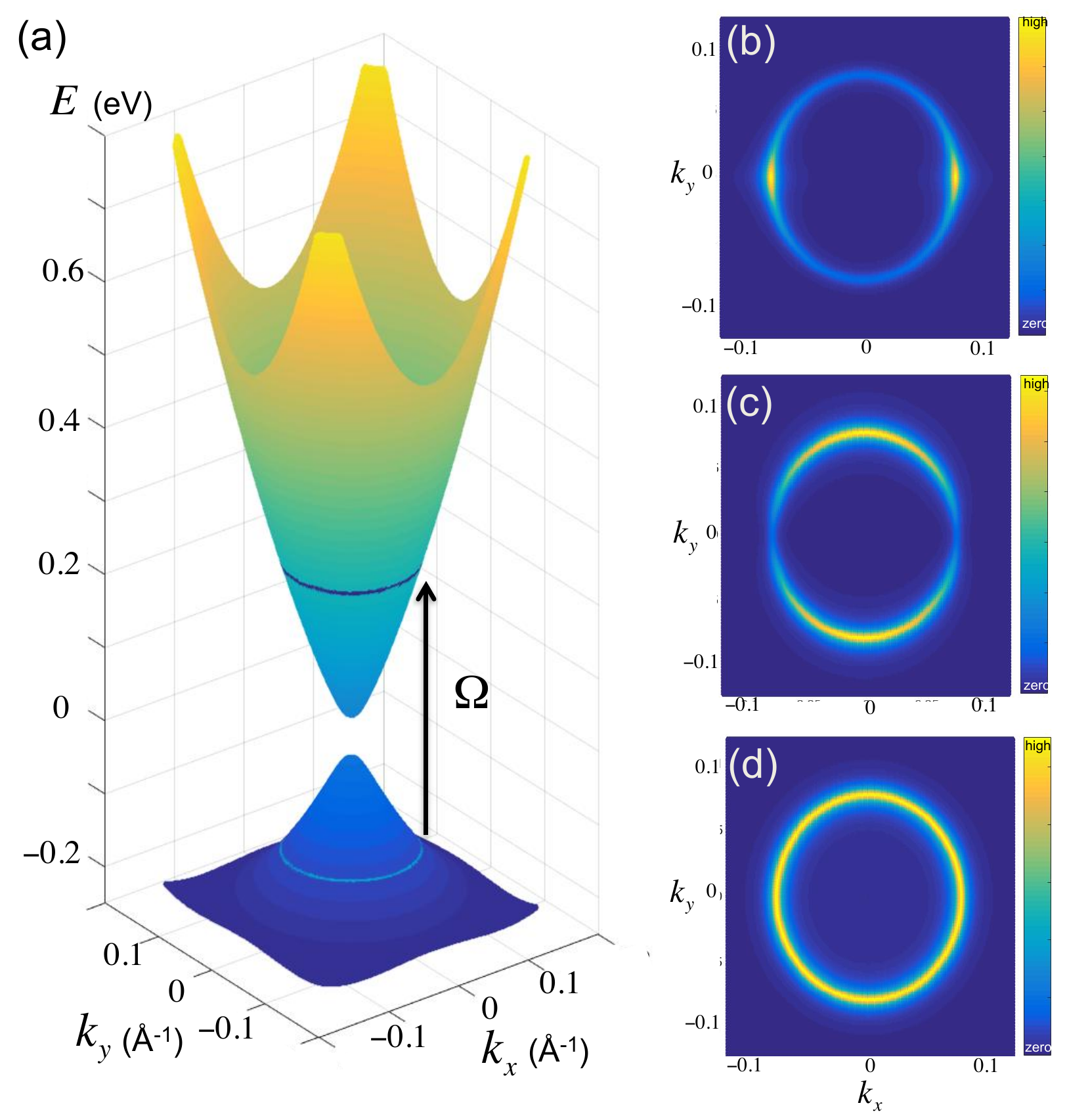}
\caption{
The susceptibility of shift spin-current, $J_{x,spin}=\chi_{x,\hat{s}_x}^{xx}E_x^2$, is decomposed into the parts of perturbations and responses. (a) The dispersion of gapped Dirac surface states with resonance condition in distinct colors. $m=0.03(eV)$, $g=7(eV\AA^2)/\hbar^2$, $v_F=2.55(eV\AA)/\hbar$ are used.  (b) The coupling strength of $\ket{\vec{k}}$  to a spin-diamagnetic current response, $\delta(\om_{cv}-\Omega)|(\{ \hat{s}_x, \partial_{k_x}v^x \})_{cv}|$, is plotted. (c) The coupling strength of  $\ket{\vec{k}}$ to an external electric field, $\delta(\om_{cv}-\Omega)|(v^x)_{vc}|$, is plotted. (d) The contribution of shift spin-current from  $\ket{\vec{k}}$, $\delta(\om_{cv}-\Omega)|(v^x)_{vc}(\{ \hat{s}_x, \partial_{k_x}v^x\})_{cv}|$, is plotted. 
}
\end{center}
\end{figure}

\beq
 v^i(\vec{k}) _{vc} (\partial_{k_j} v^l(\vec{k}))_{cv} &=& (\partial_{k_i} H(\vec{k}) )_{vc} (\partial_{k_j} \partial_{k_l} H(\vec{k}) )_{cv},\nonumber \\
&\xrightarrow{\mathcal{T} }&  (-\partial_{k_i} H(-\vec{k}) )^*_{vc} (\partial_{k_j} \partial_{k_l} H(-\vec{k}) )^*_{cv},\nonumber \\
&=& -v^i(-\vec{k})_{vc}^* (\partial_{k_j} v^l(-\vec{k}))_{cv}^*. 
\eeq
where the time-reversal operation is meant to be $H(\vec{k}) = e^{-ikr}\hat{\mathcal{H}}e^{ikr} \xrightarrow{\mathcal{T} }  e^{-ikr}\Theta^{-1}\hat{\mathcal{H}} \Theta e^{ikr}=\sigma^y H^*(-\vec{k}) \sigma_y $ (see supplementary information). Combined with the spin operator, which is to pick a diamagnetic current along a particular spin direction: 

\beq
v^i_{vc}(\vec{k}) (\{\hat{s}_n, \partial_{k_j} v^l(\vec{k}) \})_{cv} &\xrightarrow{\mathcal{T} }&  
 (v^i(-\vec{k}))_{vc}^*( \{ \hat{s}_n, \partial_{k_j} v^l(-\vec{k})\})_{cv}^* .\nonumber
\eeq
As before, the summation over $\vec{k}$ can be done for $\ket{\vec{k}}$ and $\ket{-\vec{k}}$. Taking the imaginary part of the integrand leads to the zero shift spin-susceptibility:

\beq
\chi_{j,spin}^{il}&=& \frac{1}{2}\int \frac{d^d\vec{k}}{(2\pi)^d} \delta (\om_{cv}-\Omega) \Im \bigg[  v_{vc}^i(\vec{k} )   
(\{ \hat{s}_n, \partial_{k_j} v^l (\vec{k}) \})_{cv} \nonumber \\
&&+  v_{vc}^{i} (-\vec{k} ) (\{\hat{s}_n, \partial_{k_j} v^{l} (-\vec{k}) \} )_{cv}\bigg] =0.
\eeq
This is the case for a linearly polarized light considered in this study, which preserves the time-reversal symmetry.  A simple example of a Dirac surface state with perturbations that generate a shift spin-current with zero charge-current is following:

\beq
H &=& -v_Fk_y \sigma_x + v_Fk_x \sigma_y +m \sigma_z + g(k_x^2+k_y^2) \sigma_0.\label{Hm}
\eeq
The third term on the right side is from the magnetic ordering on the surface of 3d topological insulator that breaks the time-reversal symmetry. The last term is a band-bending term that provides diamagnetic terms ($\sim A_i A_l$). Computing the susceptibilities for different combinations of indices, we obtain the following non-zero components:

\beq
-\chi_{x,\hat{s}_x}^{xx}=\chi_{x,\hat{s}_y}^{yx}=-\chi_{y,\hat{s}_x}^{xy}=\chi_{y,\hat{s}_y}^{yy}=\frac{gm}{2v_F\Omega^2}
\label{rashsus}
\eeq
for the frequency of photon larger than the energy gap, $\Omega>m$. As opposed to the warped Dirac surface state example where we only keep the leading order of warping strength $\lambda$, Eq.\eqref{rashsus} is exact results for Hamiltonian Eq.\eqref{Hm}. Note that the spin direction of photocurrent rotate with the direction of magnetic ordering, $m\sigma_z$. Together, a linearly polarized light, $\vec{E}(t) = E_0 (\cos \phi \hat{x} + \sin \phi \hat{y})\cos\Omega t$, induces the following shift spin-current:

\beq
\vec{J}_{spin}&=& 
 - \frac{gm E_0^2}{2v_F\Omega^2}  (\cos \phi \hat{x} + \sin \phi \hat{y})(\cos \phi \hat{s}_x + 
\sin \phi \hat{s}_y) \nonumber \\
\eeq 

The band-bending near the surface of topological insulators have been reported in several studies~\cite{hsie2009tuna, bian2010coex, king2011larg}. By introducing a magnetic dopants providing Zeeman term, Hamiltonian \eqref{Hm} is readily realized. In contrast to the 
shift charge-current in warped Dirac surface states, we have more flexibility to control the direction of magnetic ordering and its magnitude. With reasonable choices of $m=30 (meV)$, $\hbar \Omega = 60 (meV)$ and experimentally observed Fermi velocity and band bending parameters~\cite{king2011larg}, we obtain shift spin-current density $0.21I_0 (nA/m)$, which is in the same order with the shift charge-current estimated in the previous section. 

On the other hand, a large band-bending at the surface of topological insulators induces a 2d electron gas (2DEG) with Rashba spin splitting. The 2DEG is described by the same Hamiltonian \eqref{Hm} 
with different Fermi velocity $v_F=0.36\hbar^{-1} (eV\AA)$, one order of magnitude less than that for Dirac surface state (see supplementary information).  As a result, we obtain the 2DEG 
spin-shift current density $1.5 I_0 (nA/m)$, which is even larger than that of massive Dirac surface states. By carefully tuning the Fermi energy, one will be able to see different involvement of Dirac 
surface states and 2DEG in the production of shift spin-currents.

In summary, starting from the second order photocurrent derivation in the Floquet formalism, shift charge- and spin-currents are explained as two point correlation functions of para- and dia-magnetic currents. Two examples, both from the Dirac surface states, are shown and their theoretical values in experiments are quantified.

This work was supported by the EPiQS initiative of the Gordon and Betty
Moore Foundation (TM) and by JSPS Grant-in-Aid for Scientific Research (No. 24224009, and No. 26103006) from MEXT, Japan, and ImPACT Program of Council for Science, Technology and Innovation (Cabinet office, Government of Japan) (NN).

\bibliography{kun_biblio}

\hfill \break
\hfill \break
\hfill \break
\hfill \break

\onecolumngrid
\hfill \break

\hrulefill
\begin{center}
\textbf{\Large Supplementary Materials}
\end{center}
\hfill \break

\begin{center}
\textbf{\large S1. Shift currents and symmetries }
\end{center}

In the main text, we claim that a shift charge-current is allowed in the absence of the inversion symmetry, and a shift spin-current requires, in addition, the absence of the time-reversal symmetry. In this supplementary, we provide the details of proofs.

\subsection{Inversion symmetry and shift charge-currents}
First of all, consider a system with the inversion symmetry. The inversion symmetric Hamiltonian satisfies $[ \hat{\mathcal{H}} ,  {\mathcal{P}} ]=0$, which connects Bloch Hamiltonian $H(\vec{k})$ and $H(-\vec{k})$: 

\beq
H(\vec{k}) &=& e^{-i\vec{k}\cdot \vec{r}} \hat{\mathcal{H}} e^{i\vec{k}\cdot \vec{r}}=e^{-i\vec{k}\cdot \vec{r}} \mathcal{P}^{-1}\hat{\mathcal{H}} \mathcal{P} e^{i\vec{k}\cdot \vec{r}}\\
&=&\mathcal{P}^{-1} e^{i\vec{k}\cdot \vec{r}} \hat{\mathcal{H}} e^{-i\vec{k}\cdot \vec{r}} \mathcal{P}= \mathcal{P}^{-1} H(-\vec{k}) \mathcal{P}
\eeq
According to this, we can establish the relation of Fermi velocity between inversion symmetry partners:
\beq
v^i_{vc}(\vec{k}) &=& \bra{u_v(\vec{k})} \frac{\partial H(\vec{k})}{\partial k_i} \ket{u_c(\vec{k})} = \bra{u_v(\vec{k})} \frac{ H(\vec{k}+\Delta k \hat{i})- H(\vec{k})}{\Delta k} \ket{u_c(\vec{k})} \\
&=& -\bra{u_v(\vec{k})} \mathcal{P}^{-1} \frac{ H(-\vec{k}-\Delta k \hat{i})- H(-\vec{k}) }{-\Delta k} \mathcal{P} \ket{u_c(\vec{k})} = -\bra{u_v(\vec{k})}\mathcal{P}^{-1} \frac{\partial H(-\vec{k})}{\partial k_i} \mathcal{P}\ket{u_c(\vec{k})} \\
&=& -v_{vc}(-\vec{k})
\eeq
where in the last equality, we used the fact that $\mathcal{P}\ket{u_{c,v}(\vec{k})}$ is an eigenstate of $H(-\vec{k})$. Similarly, it is straightforward to show $(\partial_{k_j} v^l (\vec{k}))_{cv} = (\partial_{k_j} v^l (-\vec{k}))_{cv}$:
\beq
(\partial_{k_j} v^l (\vec{k}))_{cv} &=& \bra{u_c(\vec{k})} \frac{\partial^2 H(\vec{k})}{\partial k_j \partial k_l } \ket{u_v(\vec{k})}  \\
&=& \bra{u_c(\vec{k})} \mathcal{P}^{-1} \frac{\partial^2 H(-\vec{k})}{\partial k_j \partial k_l } \mathcal{P} \ket{u_v(\vec{k})} = (\partial_{k_j} v^l (-\vec{k}))_{cv}
\eeq
Therefore, the contribution to the shift charge-current from a Bloch state $\ket{\vec{k}}$ is exactly cancelled by its inversion symmetry partner $\mathcal{P}\ket{\vec{k}}$: 
\beq
v_{vc} (\vec{k}) (\partial_{k_j} v^l(\vec{k}) )_{cv} = - v_{vc} (-\vec{k}) (\partial_{k_j} v^l(-\vec{k}) )_{cv} 
\eeq
Therefore, 

\beq
\chi_j^{il}&=& \frac{1}{2}\int d\vec{k} \delta (d_z) \Im 
\bigg[ v_{vc}^i(\vec{k} )\partial_{k_j} v_{cv}^l (\vec{k})
\nonumber  + v_{vc}^i(-\vec{k} )\partial_{-k_j} v_{cv}^l (-\vec{k})\bigg] =0
\eeq
Let us comment that the inversion symmetry operation on shfit spin-current. Because $\mathcal{I}: \hat{s}_n \rightarrow \hat{s}_n$,  
\beq
(\{ \hat{s}_n, \partial_{k_j} v^l (\vec{k}) \})_{vc} &=& \bra{u_v(\vec{k})} \bigg\{ \hat{s}_n, \frac{\partial^2 H(\vec{k})}{\partial k_j\partial k_l} \bigg\} \ket{u_c(\vec{k})} \\
&=& \bra{u_v(\vec{k})}\mathcal{P}^{-1} \bigg\{ \hat{s}_n, \frac{\partial^2 H(-\vec{k})}{\partial k_j \partial k_l} \bigg\} \mathcal{P} \ket{u_c(\vec{k})} \\
&=& (\{ \hat{s}_n, \partial_{k_j} v^l (-\vec{k}) \})_{vc} 
\eeq
Therefore,
\beq
v(\vec{k})_{vc}  (\{\hat{s}_n, \partial_{k_j} v^l(\vec{k}) \} )_{cv} &=& - v(-\vec{k})_{vc}  (\{\hat{s}_n, \partial_{k_j} v^l(-\vec{k}) \} )_{cv} 
\eeq
Thus, the presence of inversion symmetry gives the exactly opposite shift spin-current contributions from its inversion symmetry partners, and the inversion symmetry must be broken for a system to carry a shift spin-current as well.

\subsection{The time-reversal symmetry and shift currents}

Then, let us consider a system with the time-reversal symmetry, $[ \Theta, \hat{\mathcal{H}}]=0$. The time reversal operation can be written $\Theta = i\sigma_y K$, where operator $K$ is complex conjugation. The Bloch wavefunction is transformed in the following way:

\beq
H(\vec{k}) &=& e^{-i\vec{k}\cdot\vec{r}} \hat{\mathcal{H}} e^{i\vec{k}\cdot\vec{r}}=e^{-i\vec{k}\cdot\vec{r}} \Theta^{-1}\hat{\mathcal{H}} \Theta e^{i\vec{k}\cdot\vec{r}}\\
&=&i\sigma_y K e^{i\vec{k}\cdot\vec{r}} \hat{\mathcal{H}} e^{-i\vec{k}\cdot\vec{r}} (-i)\sigma_y K \\
&=& \sigma_y  e^{-i\vec{k}\cdot\vec{r}} \hat{\mathcal{H}}^* e^{i\vec{k}\cdot\vec{r}} \sigma_y  \\
&=& \sigma_y  \left( e^{i\vec{k}\cdot\vec{r}} \hat{\mathcal{H}} e^{-i\vec{k}\cdot\vec{r}} \right)^* \sigma_y  \\
&=& \sigma_y  H^*(-\vec{k}) \sigma_y  \\
\eeq
Accordingly, Fermi velocity component is transformed:
\beq
v^i_{vc}(\vec{k}) &=& \bra{u_v(\vec{k})} \frac{\partial H(\vec{k})}{\partial k_i} \ket{u_c(\vec{k})} = \bra{u_v(\vec{k})} \frac{ H(\vec{k}+\Delta k \hat{i})- H(\vec{k})}{\Delta k} \ket{u_c(\vec{k})} \\
&=& -\bra{u_v(\vec{k})} \sigma_y \frac{ H^*(-\vec{k}-\Delta k \hat{i})- H^*(-\vec{k}) }{-\Delta k} \sigma_y \ket{u_c(\vec{k})} \\
&=&  -\bra{u_v(\vec{k})} \sigma_y \frac{\partial H^*(-\vec{k})}{\partial k_i} \sigma_y \ket{u_c(\vec{k})} \\
&=&-\left( ^*\bra{u_v(\vec{k})} \sigma_y \frac{\partial H(-\vec{k})}{\partial k_i} \sigma_y \ket{u_c(\vec{k})}^* \right) ^*\\
&=&-\left( \bra{u_v(\vec{k})} \Theta^{-1}\frac{\partial H(-\vec{k})}{\partial k_i}  \Theta \ket{u_c(\vec{k})} \right) ^*\\
&=& -v_{vc}^*(-\vec{k})
\eeq
where we use the fact that the time reversal partner of a state $\ket{\vec{k}}$ is an eigen state at ${-\vec{k}}$. Similarly, one can prove that 

\beq
(\partial_{k_j} v^l (\vec{k}))_{cv} &=& \bra{u_c(\vec{k})} \frac{\partial^2 H(\vec{k})}{\partial k_j \partial k_l } \ket{u_v(\vec{k})}  \\
&=& \bra{u_c(\vec{k})} \sigma_y \frac{\partial^2 H^*(-\vec{k})}{\partial k_j \partial k_l } \sigma_y \ket{u_v(\vec{k})} = (\partial_{k_j} v^l (-\vec{k}))^*_{cv}
\eeq
This proves the claim in the main body.

\begin{center}
\textbf{\large S2. Spherical coordinate expressions}
\end{center}

Spherical coordinate is  convenient to deal with resonant conditions in Dirac Hamiltonians. It converts the momentum integration to spherical angular integation with a fixed radius. Consider a Hamiltonian:
\beq
H &=& (h_1,h_2,h_3)\cdot \vec{\sigma}, \\
&=& h (\sin\theta \cos\phi, \sin\theta \sin\phi, \cos\theta) \cdot \vec{\sigma}. 
\eeq
Its eigenvectors are: 
\beq
\ket{u_c}&=&\bpm e^{-i\phi}\cos \theta/2  \\ \sin\theta/2 \epm, \,\,
\ket{u_v}=\bpm e^{-i\phi}\sin \theta/2  \\ -\cos\theta/2 \epm.
\eeq
In this gauge choice, let us compute off-diagonal components of pauli matrices:
\beq
\dirac{u_c}{\sigma_1| u_v}&=&-\cos\phi \cos\theta - i \sin\phi \\
\dirac{u_c}{\sigma_2|u_v}&=&-\sin\phi \cos\theta + i \cos \phi \\
\dirac{u_c}{\sigma_3|u_v}&=&\sin \theta
\eeq
These quantities are guage dependent, but the expression of shift current is guage-independent. A Jacobian appears as the momentum integration is converted into the spherical coordinate integration, 
\beq
\int dk_x dk_y dh_ = \int \left| \frac{\partial (k_x,k_y,h_3)}{\partial (h,\theta,\phi)} \right| dh d\theta d\phi
\eeq
In the following examples, the Jacobian is simply $J = h^2\sin\theta $.

\begin{center}
\textbf{\large S3. Shift charge-current in warped Dirac suface states}
\end{center}

Let us compute shift charge-currents in warped Dirac surface states. 
First of all, the Hamiltonian and its derivatives are (in this section, $v_F=1$ is assumed for simplicity): 
\beq
{h} &=& [-k_y, k_x, \lambda (k_x^3- 3k_x k_y^2)]\cdot \vec{\sigma},\\
{v}^x &=& \partial_{k_x} h = [ 0,1,3 \lambda (k_x^2-k_y^2)]\cdot \vec{\sigma}, \,\,\,\,
 {v}^y = \partial_{k_y} h =  [ -1,0,-6 \lambda k_x k_y]\cdot \vec{\sigma}, \\
 \partial_{k_x} v^x &=& 6\lambda k_x \sigma_3, \,\,\,\,  \partial_{k_y} v^x = -6\lambda k_y \sigma_3, \,\,\,\,  \partial_{k_x} v^y = -6\lambda k_y \sigma_3, \,\,\,\,  \partial_{k_y} v^y = -6\lambda k_x \sigma_3.
\eeq
Our goal is to compute the susceptibility of shift current, $J_j=\chi_j^{il}E_iE_l$: 
\beq
\chi_j^{il} = \frac{\pi}{\Omega^2}\int \frac{d^2\vec{k}}{4\pi^2} \delta(d_z)  (v^i)_{vc} (\partial_{k_j} v^l)_{cv} 
\eeq
Using spherical coordinates: 
\beq
h_1 &=&h \sin\theta \cos\phi = -k_y \\
h_2 &=& h \sin\theta \sin\phi = k_x \\
h_3 &=& h \cos\theta = \lambda (k_x^3- 3k_x k_y^2) = \lambda h^3 \sin^3\theta \cos\phi (\cos^2\phi  - 3 \sin^2\phi )
\eeq
The 2d momentum integration is converted to spherical coordinate integration with a delta function: 
\beq
\int dk_x dk_y &=& \int dk_xdk_ydh_3 \delta(h_3-h\cos\theta)\\
&=&\int \left| \frac{\partial (k_x,k_y,h_3)}{\partial (h,\theta,\phi)} \right| dh d\theta d\phi \frac{1}{h} \delta (\cos\theta - \lambda h^2 \sin^3\theta \cos\phi (\cos^2\phi  - 3 \sin^2\phi ))
\eeq
where the Jacobian is $ \left| \frac{\partial (k_x,k_y,h_3)}{\partial (h,\theta,\phi)} \right|  = h^2 \sin\theta$. Let us compute $\chi_y^{yy}$ explicitly: 

\beq
\chi_y^{yy} &=& \frac{\pi}{\Omega^2}\int \frac{d^2\vec{k}}{4\pi^2} \delta(\om_{cv}-\Omega) \Im[ (v^y)_{vc} (\partial_{k_y} v^y)_{cv} ], \\
&=&\frac{1}{4\pi\Omega^2} \int dh d\theta d\phi h^2 \sin\theta  \delta(2h-\Omega)  \frac{1}{h} \delta (\cos\theta +O(\lambda)) \Im[ \bra{u_v}-\sigma_1 +O(\lambda) \ket{u_c} \bra{u_c} -6\lambda k_x \sigma_3 \ket{u_v}] \\
&\simeq& \frac{1}{4\pi\Omega^2} \int dh d\theta d\phi h^2 \sin\theta  \delta(2h-\Omega)  \frac{1}{h} \delta (\cos\theta ) \Im[ \bra{u_v}-\sigma_1 \ket{u_c} \bra{u_c} 6\lambda h\sin\theta \sin\phi \sigma_3 \ket{u_v}] \\
&=&  \frac{1}{4\pi\Omega^2} \int dh d\theta d\phi h^2 \sin\theta \delta(2h-\Omega)  \frac{ -6\lambda h\sin\theta \sin\phi }{h} \delta (\cos\theta )\Im[ \bra{u_v}\sigma_1 \ket{u_c} \bra{u_c}\sigma_3 \ket{u_v} ]\\
&=&  \frac{1}{4\pi\Omega^2} \int dh d\theta d\phi h^2 \sin\theta \frac{\delta(h-\Omega/2)}{2}  \frac{ -6\lambda h\sin\theta \sin\phi }{h} \delta (\cos\theta ) \Im[(-\cos\phi \cos\theta + i\sin\phi) \sin\theta ]\\
&=& -\frac{1}{4\pi\Omega^2} \int d\phi \left( \frac{\Omega}{2}\right)^2  3\lambda  \sin^2\phi   \\
&=&  -\frac{3\lambda }{16},
\eeq
where the leading order of $\lambda$ is kept only. This exactly reproduces the result in the main text with $v_F=1$.   Similarly, other components can be computed in a straightforward manner: 

\beq
\chi_x^{yx} &=& \frac{\pi}{\Omega^2}\int \frac{d^2\vec{k}}{4\pi^2} \delta(\om_{cv}-\Omega)  \Im[(v^y)_{vc} (\partial_{k_x} v^x)_{cv}], \\
&\simeq& \frac{1}{4\pi\Omega^2} \int dh d\theta d\phi h^2 \sin\theta  \delta(2h-\Omega)  \frac{1}{h} \delta (\cos\theta ) \Im[ \bra{u_v}-\sigma_1 \ket{u_c} \bra{u_c} -6\lambda h\sin\theta \sin\phi \sigma_3 \ket{u_v}] = \frac{3\lambda }{16}, \\
\chi_x^{xy} &=& \frac{\pi}{\Omega^2}\int \frac{d^2\vec{k}}{4\pi^2} \delta(\om_{cv}-\Omega)  \Im[ (v^x)_{vc} (\partial_{k_x} v^y)_{cv} ], \\
&\simeq& \frac{1}{4\pi\Omega^2} \int dh d\theta d\phi h^2 \sin\theta  \delta(2h-\Omega)  \frac{1}{h} \delta (\cos\theta )\Im[  \bra{u_v}\sigma_2 \ket{u_c} \bra{u_c} 6\lambda h\sin\theta \cos\phi \sigma_3 \ket{u_v} ]= \frac{3\lambda }{16}, \\
\chi_y^{xx} &=& \frac{\pi}{\Omega^2}\int \frac{d^2\vec{k}}{4\pi^2} \delta(\om_{cv}-\Omega) \Im[ (v^x)_{vc} (\partial_{k_y} v^x)_{cv}], \\
&\simeq& \frac{1}{4\pi\Omega^2} \int dh d\theta d\phi h^2 \sin\theta  \delta(2h-\Omega)  \frac{1}{h} \delta (\cos\theta ) \Im[ \bra{u_v}\sigma_2 \ket{u_c} \bra{u_c} 6\lambda h\sin\theta \cos\phi \sigma_3 \ket{u_v} ]= \frac{3\lambda }{16}.
\eeq

\begin{center}
\textbf{\large S4. Shift spin-current in massive Dirac surface states}
\end{center}

By introducing magnetic ordering at the surface of 3d topological insulator, the time-reversal symmetry is broken and we find shift spin-currents linearly proportional to band bending parameter. The Hamiltonian and its derivatives are: 

\beq
{h} &=& [-k_y, k_x, m ]\cdot \vec{\sigma} + g (k_x^2+k_y^2)\sigma_0 ,\\
{v}^x &=& \partial_{k_x} h = \sigma_2+2gk_x\sigma_0,\,\,\,\, {v}^y = \partial_{k_y} h =-\sigma_x+2gk_y\sigma_0,\\
\partial_{k_x} v^x &=& 2g \sigma_0, \,\,\,\, \partial_{k_y} v^y = 2g\sigma_0.
\eeq
For spin shift-current, we need pick a particular spin direction in the diamagnetic current: 
\beq
\{ \hat{s}_n, \partial_{k_j}v^l \} = \frac{1}{2}[ \sigma_n (\partial_{k_j}v^l )+(\partial_{k_j} v^l) \sigma_n]
\eeq
Our goal is to compute the susceptibility of shift spin-current, $J_{j,\hat{s}_n}=\chi_{j,\hat{s}_n}^{il} E_i E_l$: 
\beq
\chi_{j,\hat{s}_n}^{il} &=& \frac{\pi}{\Omega^2}\int \frac{d^2\vec{k} }{(2\pi)^2}\delta(\om_{cv}-\Omega)  \Im \big[ (v^i)_{vc} (\partial_{k_j} \{\hat{s}_n, v^l \})_{cv} \big] ,
\eeq
The integration over momentum is converted to spherical coordinate integration with a delta funtion: 
\beq
\int dk_x dk_y  &=& \int dk_xdk_ydh_3 \delta(h_3-h\cos\theta)\\
&=&\int  \left| \frac{\partial (k_x,k_y,h_3)}{\partial (h,\theta,\phi)} \right| dh d\theta d\phi \frac{1}{h} \delta \left( \cos\theta - \frac{m}{h} \right)
\eeq
 where the Jacobian is $ \left| \frac{\partial (k_x,k_y,h_3)}{\partial (h,\theta,\phi)} \right|  = h^2 \sin\theta$. For example, let us get  susceptibility $\chi_{x,\hat{s}_x}^{xx}$: 
 \beq
 \chi_{x,\hat{s}_x}^{xx}
 &=& \frac{\pi}{\Omega^2}\int \frac{d^2\vec{k}}{4\pi^2} \delta(\om_{cv}-\Omega) \Im \big[  (v^x)_{vc} (\partial_{k_x} \{\hat{s}_x, v^x \})_{cv} \big] ,\\
 &=&\frac{1}{4\pi\Omega^2} \int dh d\theta d\phi h^2\sin\theta \frac{1}{h} \delta \left( \cos\theta - \frac{h_3}{h} \right) \frac{\delta(h-\Omega/2)}{2}\Im [ \bra{u_v} \sigma_y + 2k_xs_0 \ket{u_c} \bra{u_v} 2g \sigma_x \ket{u_c}] \\
  &=&\frac{1}{4\pi\Omega^2} \int dh d\theta d\phi h\sin\theta  \delta \left( \cos\theta - \frac{h_3}{h} \right)\frac{\delta(h-\Omega/2)}{2} \Im [(-\sin\phi \cos\theta - i \cos\phi )2g (-\cos\phi\cos\theta - i \sin\phi )] \\
&=&\frac{1}{4\pi\Omega^2} \int dh d\theta d\phi h\sin\theta  \delta \left( \cos\theta - \frac{h_3}{h} \right) \frac{\delta(h-\Omega/2)}{2} 2g \cos\theta \\
&=&- \frac{gm}{2\Omega^2}
 \eeq
Note that there is no approximation taken in this calculation. Therefore, as long as the continuum Hamiltonian is valid, the shift spin-current expression obtaind here is exact. Other components can be calculated similarly: 
\beq
 \chi_{x,\hat{s}_y}^{yx}&=& \frac{\pi}{\Omega^2}\int \frac{d^2\vec{k}}{4\pi^2} \delta(\om_{cv}-\Omega)  \Im \big[ (v^y)_{vc} (\partial_{k_x} \{\hat{s}_y, v^x \})_{cv} \big] ,\\
 &=&\frac{1}{4\pi\Omega^2} \int dh d\theta d\phi h^2\sin\theta \frac{1}{h} \delta \left( \cos\theta - \frac{h_3}{h} \right) \frac{\delta(h-\Omega/2)}{2}\Im [ \bra{u_v} -\sigma_x  \ket{u_c} \bra{u_v} 2g \sigma_y \ket{u_c}] = - \frac{gm}{2\Omega^2},\\
  \chi_{y,\hat{s}_x}^{xy}&=& \frac{\pi}{\Omega^2}\int \frac{d^2\vec{k}}{4\pi^2} \delta(\om_{cv}-\Omega)  \Im \big[  (v^x)_{vc} (\partial_{k_y} \{\hat{s}_x, v^y \})_{cv} \big] ,\\
 &=&\frac{1}{4\pi\Omega^2} \int dh d\theta d\phi h^2\sin\theta \frac{1}{h} \delta \left( \cos\theta - \frac{h_3}{h} \right) \frac{\delta(h-\Omega/2)}{2}\Im [ \bra{u_v} \sigma_y  \ket{u_c} \bra{u_v} 2g \sigma_x \ket{u_c}] =  \frac{gm}{2\Omega^2}, \\
   \chi_{y,\hat{s}_y}^{yy}&=& \frac{\pi}{\Omega^2}\int \frac{d^2\vec{k}}{4\pi^2} \delta(\om_{cv}-\Omega)  \Im \big[  (v^y)_{vc} (\partial_{k_y} \{\hat{s}_y, v^y \})_{cv} \big] ,\\
 &=&\frac{1}{4\pi\Omega^2} \int dh d\theta d\phi h^2\sin\theta \frac{1}{h} \delta \left( \cos\theta - \frac{h_3}{h} \right) \frac{\delta(h-\Omega/2)}{2}\Im [ \bra{u_v} -\sigma_x  \ket{u_c} \bra{u_v} 2g \sigma_y \ket{u_c}] =  -\frac{gm}{2\Omega^2}, \\
 \eeq
Other combinations of indices yield zero susceptbility.

\hfill \break
\begin{center}
\textbf{\large S5. Numerical estimation of shift currents}
\end{center}
\hfill \break

In this section, we  show the numerical estimation of shift charge- and spin-current for the systems discussed in the main text. We restore the fundamental constants ($e$ and $\hbar$) back to the current expressions, and obtained expected numerical values in experiments.

\subsection{Shift charge-current in warped Dirac surface states}

The warped Dirac Hamiltonian is:

\beq
H &=& -v_F \hbar k_y \sigma_x + v_F \hbar k_x \sigma_y +\lambda(k_x^3-3 k_x k_y^2)\sigma_z  \\
 &=& -v_F p_y \sigma_x + v_F p_x \sigma_y +\frac{\lambda}{\hbar^3}(p_x^3-3 p_x p_y^2)\sigma_z 
\eeq
where $p_j= \hbar k_j$. We will use $v_F= 2.55 (eV \AA)=3.87\times 10^{5} (m/s)$ and $\lambda = 250 (eV \AA^3)$ following the estimation done by Fu~\cite{fu2009}.  Current operators are:
\beq
\hat{j}^i_{para} &=& \left. \frac{\partial H}{\partial A_i}\right|_{\vec{A}=0} = -e \frac{\partial H}{\partial p_i} =-\frac{e}{\hbar} \frac{\partial H}{\partial k_i} = -\frac{e}{\hbar} v^i \\
\hat{j}^l_{j,dia} &=& \left. \frac{\partial^2 H}{\partial A_l \partial A_j}\right|_{\vec{A}=0} A^l= e^2 \frac{\partial^2H}{\partial p_l \partial p_j} A^l =\frac{e^2}{\hbar^2} \frac{\partial^2 H}{\partial k_l \partial k_j} A^l  = \frac{e^2}{\hbar^2} (\partial_{k_j} v^l ) A^l 
\eeq
The susceptibility is:
\beq
\chi_{j}^{il} =\frac{1}{\hbar^2 \Omega^2}\int d^2\vec{p} d(\hbar \om) Tr [ \hat{j}^i_{para} G(\om+\Omega,\vec{p}) (\hat{j}^l_{j,dia} /A^{l})G(\om,\vec{p})] \nonumber \label{suscep} \\
\eeq 
And, the shift current density is:
\beq
J_j^{il} =\frac{E^i E^l }{\hbar^2 \Omega^2}\int d^2\vec{p} d(\hbar \om) Tr [ \hat{j}^i_{para} G(\om+\Omega,\vec{p}) (\hat{j}^l_{j,dia}/A^l )G(\om,\vec{p})] \nonumber \label{suscep} \\
\eeq 
As a result, we have the shift current density:
\beq
J_{j,warp}^{il} &=& \left( \frac{3 e^3 \lambda}{16 \hbar^3 v_F^2} \right) E^i E^l = \left( \frac{3 e^3 \lambda}{16 \hbar^3 v_F^2} \right)  \frac{2 I_0}{\epsilon_0 c} \simeq 0.13 I_0 (nA/m)
\eeq
where $I_0$ is intensity of light, $\epsilon_0$ is permitivity, $c$ is the speed of light. To make a comparison with experiments, it is convenient to think about current per unit intensity. And, the evaluation of this is dependent of beam size shined on samples. Let us take diameter of the beam size $l=1(mm)$ which is used in recent experiment by Okada et al.~\cite{okad2016} and also in the review article by Ganichev and Prettl~\cite{gani2003spin} (page 956) for the maximally focused beam size. Thus, shift current per intensity for size $1(mm)$ is:

\beq
\frac{I_{j,warp}^{il}}{I_0} = \frac{J_{j,warp}^{il} l}{I_0}= 0.13 (nA W^{-1}\cdot m) \times 1 (mm)= 1.3 (nAW^{-1} cm^2)
\eeq
where we take the unit $(AW^{-1}cm^2)$ following Okada et al.~\cite{okad2016}. In their experiment, they obtained photocurrent in the order of 10$(pAW^{-1}cm^2)$ (page 5 of their arxiv preprint). Our value is two orders of magnitude larger. It is possible that the assumption, $\text{photocurrent}=(\text{current-density})\times(\text{beam-size})$, is overestimating actual current measured through leads.

On the other hand, compared to Lindner et al.'s work~\cite{lind2014}, where they have a photocurrent density $10^{-8} (A/m)$ with sunlight ($I_0 \simeq 140 (W/m^2)$ as a modest estimation of sunlight intensity on Earth), our current density value $J_{j,warp}^{il} \simeq 0.13 \times 140 (nA/m) = 1.8 \times 10^{-8} (nA/m)$ is in the same order. While they introduced a magnetic strip patterning on TI surface to break both the time-reversal symmetry and in-plane rotational symmetry, we only require warping effect which is naturally presenet in discovered topological insulators.

Also, our value is also very comparble with the theoretical estimation made by Hosur in circular photo-galvanic effect. His estimation is $100 (nA/mm)$ with a $1(W)$ laser for $10(T)$ of in-plane magnetic field. If we assume that the size of sample is $1(mm^2)$, the intensity of light is $I_0 = 1(W)/1(mm^2)=10^6 (W/m^2)$. Then, our photocurrent density is $0.13 I_0 (nA/m) = 0.13\times 10^6 \times 10^{-3} (nA/mm)= 130 (nA/mm)$.

\subsection{Shift spin-current in gapped Dirac surface states}

Shift spin-current is carried in Dirac surface states with band bending and magnetic ordering that breaks the time-reversal symmetry. Hamiltonian is:

\beq
H &=& -v_F \hbar k_y \sigma_x +v_F \hbar  k_x \sigma_y +m \sigma_z + g\hbar^2(k_x^2+k_y^2) \sigma_0 \\
&=& -v_F p_y \sigma_x +v_F  p_x \sigma_y +m \sigma_z + g(p_x^2+p_y^2) \sigma_0
\eeq
The induced photocurrent density is:
\beq
J_{j,\hat{s}_n}^{il} = \frac{e^3 g m}{2 v_F (\hbar \Omega)^2} E_iE_l =  \frac{e^3 g m}{2 v_F (\hbar \Omega)^2} \frac{2I_0}{\epsilon_0 c} \simeq 0.31 I_0 (nA/m)
\eeq
where we use $m=30 (meV)$, $\hbar\Omega=60 (meV)$, and band bening parameter $g=0.36 eV \AA/(\hbar^2 k_c) \simeq 1.0 \times 10^{30} (kg\cdot m/s)^{-2}$. The band bending paramter is estimated from the ARPES spectrum by King et al.~\cite{king2011larg} which will be further discussed in the next section. Here, note that the induced shift spin-current is in the same order with the shift charge-current estimated in the previous section for typical choices of Zeeman coupling and photon frequency.

\subsection{Shift spin-current in 2d electron gas}

It is interesting that there are 2d electron gas confined in the surface of 3d topological insulators by a large band bending~\cite{bian2010coex, king2011larg}. King et al.~\cite{king2011larg} provided numerical values of Rashba coupling strength and effective mass. Consider a Hamiltonian:

\beq
E^\pm (k) =E_0 + \frac{\hbar^2 k^2}{2m^*} \pm \alpha k 
\eeq 
In the experiment~\cite{king2011larg}, it is observed that $\alpha=0.36 (eV \AA)$. Effective mass $m^*$ can be estimated by finding $k_c\simeq 0.05 (\AA^{-1})$ in their spectrum such that:

\beq
\frac{\hbar^2 k_c^2}{2m^*} - \alpha k_c=0 
\eeq
Thus, the band-bending parameter is:
\beq
g=\frac{1}{2m^*} = \frac{\alpha}{\hbar^2 k_c },
\eeq
and, $v_F= \alpha/\hbar$ in terms of parameters in Hamiltonian in the previous section. Again, assuming  Zeeman term $m=30 (meV)$ and  $\hbar \Omega=60 (meV)$, 
\beq
J_{j,\hat{s}_n}^{il}= \frac{e^3 g m}{2 v_F  (\hbar  \Omega)^2} \frac{2I_0}{\epsilon_0 c}=\frac{e^3 m }{2   k_c\hbar^3  \Omega^2} \frac{2 I_0}{\epsilon_0 c} \simeq 1.5 I_0 (nA /m)
\eeq
which is an order of magnitude larger than the shift currents induced by topological surface states.

\end{document}